# *Complex economics of simple periodic systems*

Petri P. Kärenlampi[*]

Lehtoi Research, Finland

petri.karenlampi@professori.fi

[*] Author to whom correspondence should be addressed.

**Abstract**

This paper investigates the financial economics of simple periodic systems. Well-established financial procedures appear to be complicated, and lead to partially biased results. Probability theory is applied, and the focus is on the finances of simple periodic growth processes, in the absence of intermediate divestments. The expected value of the profit rate, derived from accounting measures on an accrual basis, does not depend on the capitalization path. The expected value of capitalization is path dependent. Because of the path-dependent capitalization, the return rate on capital is path-dependent, and the time-average return rate on capital differs from the expected value of the return rate on capital for the growth cycle. The internal rate of return, defined through a compounding equation, is path-independent, thereby differing from the expected value of the rate of return on capital. It is shown that within a production estate, the area-average of internal rate of return is not representative of the rate of return on capital. The growth cycle length maximizing the return rate on equity is independent of market interest rate. Leverage effect enters the microeconomics of the growth processes through a separate leverage equation, where the leverage coefficient may reach positive or negative values. The leverage effect on the internal rate of return and the net present value are discussed. Both effects are solvable, resulting in incorrect estimates.

**Author Summary**

Economics of periodic growth systems are investigated. In such systems, no time instant is of special importance, and expected values of observables within any period are of interest. The rate of return on capital (*RROC*), on an accrual basis, differs remarkably from the internal rate of return (*IRR*) on a cash flow basis. Both differ from the net present value of cash flows (*NPV*), which further depends on the selection of the "present time". Suitable period durations do not depend on external interest rates. Leverage effects on *RROC* are successfully introduced, whereas both *IRR* and *NPV* fail to show a meaningful leverage effect.

## 1. Introduction

For more than a century, two families of criteria for capital investment decisions have coexisted: net present value (*NPV*) [1, 2, 3, 4] and internal rate of return (*IRR*) (or marginal efficiency of capital) [5, 6, 7, 8]. The first criterion, based on a discounted utility argument, apparently falls into the neoclassical realm, neglecting resource limitations of resources like the availability of capital [9, 10]. The latter, analyzing the time schedule of cash transactions, has been criticized for the existence of multiple solutions, some of them complex [11, 12, 13]. Further techniques, related to the above, exist [2, 3, 4, 14, 15].

Specifically, three different shortcomings have been identified in the net present value approach. Firstly, reinvestment of cash flows has not been included [16, 17, 11]. Procedures for overcoming this difficulty have later been developed [18, 19, 20, 21]. Secondly, the separation theorem of Fischer has been applied, resulting in path-independency of the financial criterion [2, 3, 11, 4]. Thirdly, the *NPV* approach is inherently focused on the maximization of consumption utility, instead of wealth increment [7, 22, 23, 24, 17, 25, 4, 11].

As will be detailed in the following section, the *NPV*, as well as the *IRR* are defined through compounding equations [26]. Return on investment (*ROI*), on the contrary, is defined as the ratio of operating profit to capitalization [27, 28, 29, 30, 26]. It is worth noting that internal measures of valuation not based on or used in compounding equations exist [31, 32, 33, 34]. *ROI* is not compatible with *NPV*, neither *IRR* [35]. One can ask, why they should be, considering that *NPV* approach is focused on the maximization of consumption utility, and the *IRR* on wealth increment, both based on compounding equations [7, 22, 23, 24, 17, 25, 4, 11]. Interestingly, widely varying definitions of the *ROI* appear in the literature. Magni and Marchioni [35] and Hazen and Magni [36] compute *ROI* as the ratio of the sum of discounted incomes to the sum of discounted capitalizations, as a present value at the time of an investment. Such a procedure is unconventional, but possibly justified in the case an investment project with heavy involvement at the time of initial investment. Problematically, the present values depend on the discount rate, the choice of which is inherently subjective [7, 4, 11]. It would be more conventional to discuss *ROI* within any instant of time, such metrics naturally forming a time series [37, 38, 39, 40, 41]. Then, the rate of return on investment (*RROI*) would become a time rate, integrable over time.

Even if *IRR* and *NPV* are not compatible, techniques for bridging the approaches exist [42, 43, 36]. It also is worth noting that the concept of marginal economic efficiency has been used also for another meaning [30, 44] than the marginal efficiency of capital above [7, 8]. Further, some authors have referred with *internal rate of return* to measures that are not internal to any production process but relate to market interest rates [2, 3, 4, 14].

The above-mentioned criteria for capital investment decisions lead to complex problematics. This paper intends to clarify such problematics by discussing its action on simple periodic systems. Additionally, the simple systems are discussed from the viewpoint of financial arguments simple enough to become non-disputable.

In the remaining part of this paper, the two established families of investment criteria, along with the hereby formulated rate of return on capital (*RROC)*, are applied to simple periodic growth processes. Differences in the outcomes are investigated, including the dependency of the results on the capitalization path. Finally, internal consistency of any financial approach in the presence of a leverage effect is investigated.

## 2. Methods and Results

### 2.1. Elementary financial arguments

The simple financial arguments to be applied here consist of the application of elementary probability theory to the rate of return on capital [45, 46, 47, 48]. All financial approaches are applied to simple periodic growth processes [47, 48, 49]. Even if consequent growth cycles may not be completely similar in real life, the discussion is simplified by the application of the periodic boundary condition. A further simplification for the systems discussed is the absence of intermediate divestment: the periodic growth process is not disturbed by intermediate harvest.

As the discussion regards a growth process, in the absence of intermediate harvest, the capitalization increases along with the growth process. In the determination of *ROI* or *RROI* (rate of return on investment), no single time instant is decisive. Consequently, we discuss the expected value of *RROI* within any growth period. Further, a growth process being discussed, the expected value of the capitalization is not necessarily determined as the accumulated cash flow into investment goods. That is why we now turn the discussion into the expected value of the rate of return on capital (*RROC*) instead of rate of return on investment (*RROI*). This is simply produced by computing the expected value of the operating profit rate, divided by the expected value of capitalization.

The elementary financial arguments applied here are formulated as follows. The expected value of the operating profit rate is

$$\left\langle \frac{d\kappa}{dt} \right\rangle = \int_{b}^{b+\tau} \frac{d\kappa}{dt} \; p(t) \; dt \qquad (1),$$

where $\tau$ is cycle (or period) duration, $p(t)$ is the probability density of time within the cycle, and $\frac{d\kappa}{dt}$ is any current operating profit rate. On the profit/loss – basis, the profit rate includes value growth, operative expenses, interests, and amortizations, but neglects investments and withdrawals. It is worth noting that the profit rates in Eq. (1) are additive, instead of compounding [50, 26]. On the other hand, the expected value of the capitalization is

$$\left\langle K \right\rangle = \int_{b}^{b+\tau} K \; p(t) \; dt \qquad (2),$$

where the capitalization *K*, on the balance sheet basis, is directly affected by any investment and withdrawal. Again, the capitalizations in Eq. (2) are simply additive. Then, the expected value of the rate of return on capital (*RROC*) is

$$\langle s \rangle \equiv \frac{\left\langle \frac{d\kappa}{dt} \right\rangle}{\langle K \rangle} \qquad (3).$$

As the capitalization in Eq. (2) depends on both accumulated profits and investments, at any time within the rotation cycle *b+t* the capitalization can be written

$$K(b+t) = K(b) + \int_{b}^{b+t} \frac{d\kappa}{dt} dt + \int_{b}^{b+t} \frac{dI}{dt} dt \qquad (4),$$

where $\frac{dI}{dt}$ refers to the rate of investments and divestments. The operating profit rate is written

$$\frac{d\kappa}{dt}(t) = K(t) r(t) \qquad (5).$$

In addition to the elementary financial arguments given above, the basic features of the two established families of financial criteria are given as follows. Internal rate of return *o* refers to a discount rate where discounted cash flow approaches zero, or

$$0 = \int_{b}^{b+\tau} \frac{dC}{dt} e^{-ot} dt \qquad (6),$$

where $\frac{dC}{dt}$ refers to cash flow rate. In general, Eq. (6) has many solutions for the discount rate, many of them complex [11, 12, 13]. It is worth noting that in Eq. (6), the discount rates are compound rates [50, 26]. Eq. (6) can be discretized to a polynomial, then having as many solutions as is the degree of the polynomial [12, 13]. However, most of the solutions have limited application [51, 8, 11].

The net present value of future cash flows is simply given as

$$NPV(b) = \int_{b}^{\infty} \frac{dC}{dt} e^{-qt} dt \qquad (7),$$

where q refers to discount rate. It is found that the right-hand sides of Eqs. (6) and (7) are of the same form. Interestingly, unlike Eq. (3), the value of Eqs. (6) and (7) depend on the phase of the period (or the time instant *b*) where the computation is started. Again, the discount rate in Eq. (7) is compounding.

The internal rate of return (*IRR*) according to Eq. (6) is traditionally computed on cash flow basis [7, 4, 8]. In this paper, three featuress are included in the elementary financial treatment. Firstly, the return on capital is discussed as an integrable time rate (*RROC*), according to Eq. (3). Secondly, the profit rate appearing in Eq. (1) and then substituted to Eq. (3) is discussed on an accrual basis. Thirdly, accrual-basis increments of capitalization are discussed as consequences of a growth process, contributing positively to the profit rate.

An apparently nontrivial question is, how the ratio of the additive expected values in Eq. (3) relates to the compounding discount rate in Eqs. (6) and (7). Accounting is inherently based on simply additive processes, whereas financial computations are often compounding [50, 26]. The discount rates in Eqs. (6) and (7) are defined through the compounding equations. The rate of return on capital (*RROC*) in Eq. (3) is defined through accounting equations, but it is supposed to be used in compounding equations, in the clarification of the change of wealth over time [7, 22, 23, 24, 17, 25, 11, 47]. This apparently is a remarkable difference between Eq. (3) on the one hand, and (6) and (7) on the other; the former is rigorously derived from accounting, whereas the latter ones are not.

## 2.2. Return rate on capital and internal rate of return

In periodic growth systems, profits often accumulate within a rotation period. On accrual basis, instead of cash basis, the growth of any season adds to profit. In the absence of intermediate divestments, the expected value of the operating profit rate can be written

$$\left\langle \frac{d\kappa}{dt} \right\rangle = \int_0^\tau \frac{d\kappa}{dt} \; p(t) \; dt = \int_0^\tau K(t) r(t) \; p(t) \; dt = \int_0^\tau K(0) \exp\left(\int_0^t r dt'\right) r(t) \; p(t) \; dt \qquad (8),$$

where the arbitrary starting point of the integration that appeared in Eq. (1) has been abandoned, and the time origin is placed at the beginning of the rotation cycle. The reason for this arrangement is that divestment often occurring at the end of any rotation cycle does then not become an intermediate divestment. On the other hand, the expected value of the operating profit rate can be given simply by normalizing the total accumulated operating profit by the duration of the rotation period:

$$\left\langle \frac{d\kappa}{dt} \right\rangle = \frac{1}{\tau}\left[\kappa(\tau) - \kappa(0)\right] = \frac{1}{\tau}\left[ K(0) \exp\left(\int_0^\tau r dt\right) - K(0) \right] = \frac{K(0)}{\tau}\left[ e^{\tau \langle r \rangle} - 1 \right] \qquad (9).$$

The last form of Eq. (9) shows that the expected value of the profit rate is path-independent: it depends on the time-average value of the spot return rates $\langle r \rangle$, rather than the sequence of the return rates. The path-independency does not become violated by eventual negative returns.

On the other hand, in the absence of intermediate divestments, the expected value of the capitalization can be written

$$\langle K \rangle = \int_0^\tau K \; p(t) \; dt = \int_0^\tau K(0) \exp(\int_0^t r dt') \; p(t) \; dt \qquad (10).$$

Interestingly, there is no path-independent form of Eq. (10). Correspondingly, the expected value of the capitalization is path dependent. It depends on the time schedule of the spot return rates in the inner integral of Eq. (10). This will naturally render the return rate on capital according to Eq. (3) path dependent. At elevated return rates at the beginning of the rotation period, the expected value of capitalization becomes greater, and the expected value of the return rate on capital is smaller. Reduced return rates at the beginning of the rotation have the opposite effect.

The path-dependence of the return rate on capital appears worthy of investigation. A simple *Ansatz* for the path-dependency might be to write the spot return rate on capital as a function of time on period, period duration, and the time-average value of the rate of return on capital:

$$r(t) = \langle\langle r \rangle\rangle \left[ a + (1-a) \frac{2}{\pi} \sin^2\left( \frac{t}{\Gamma} \pi \right) \right] \qquad (11),$$

where $\langle\langle r \rangle\rangle$ is the time-average value of the spot rate of return on capital within a full cycle $\frac{\Gamma}{\Gamma}\pi$ of the squared periodic function, and $a$ is a modeling parameter. *Ansatz* here refers to a tentative formulation established for the purpose of illustrating a problem.

The above discussion can be readily compared with an internal rate of return, determined on a cash-flow basis. In the absence of intermediate divestments, Eq. (6) for *IRR* can be rewritten

$$K(\tau)e^{-\sigma\tau} - K(0) = K(0)e^{<r>\tau}e^{-\sigma\tau} - K(0) = 0 \qquad (12),$$

where the time-average value of the rate of return on capital $\langle r \rangle$ depends on the rotation age $\tau$ according to the *Ansatz* of Eq. (11).

Figure 1 shows the annual spot return rate according to Eq. (11), the expected value of return on capital according to Eq. (3), and the internal rate of return according to Eq. (12) as a function of the phase of the full period cycle of duration $\pi$ from Eq. (11). In Fig. 1, the parameter value $a$=0.5. As the internal rate of return $o$ is path-independent according to Eqs. (6) and (12), it converges to the full-cycle average $\langle\langle r \rangle\rangle$ at the end of the full cycle.

Eq. (12) readily shows that the internal rate of return $o$ corresponds to the time-average of the spot return rates. On the other hand, according to Eq. (3), (8) and (10), the expected value of the return on capital is produced by weighing the spot return rates by current capitalization. In the absence of intermediate divestments, capitalization increases within a growing cycle. In Figure 1, the spot return rate, resulting from Eq. (11), is symmetric with respect to the center point of the growth period, where it displays its maximum. The low-return rate cycle part towards the end of the growth cycle is weighed by the greatest capitalization, rendering the expected rate of return $\langle s \rangle$ from Eq. (3) at the end of the full growth cycle lower than the time-average $\langle\langle r \rangle\rangle$.

The maximum value of either *IRR* ($o$) or the expected value of the rate of return on capital $\langle s \rangle$ is not reached at the end of the full period cycle of duration $\pi$ from Eq. (11), but at an earlier phase. The expected value of the *RROC* (or $\langle s \rangle$) being path-dependent, and the capitalization being smaller at early stages of the cycle, the maximum value of $\langle s \rangle$ is greater than that of *IRR* ($o$), and $\langle s \rangle$ reaches its maximum earlier within the cycle.

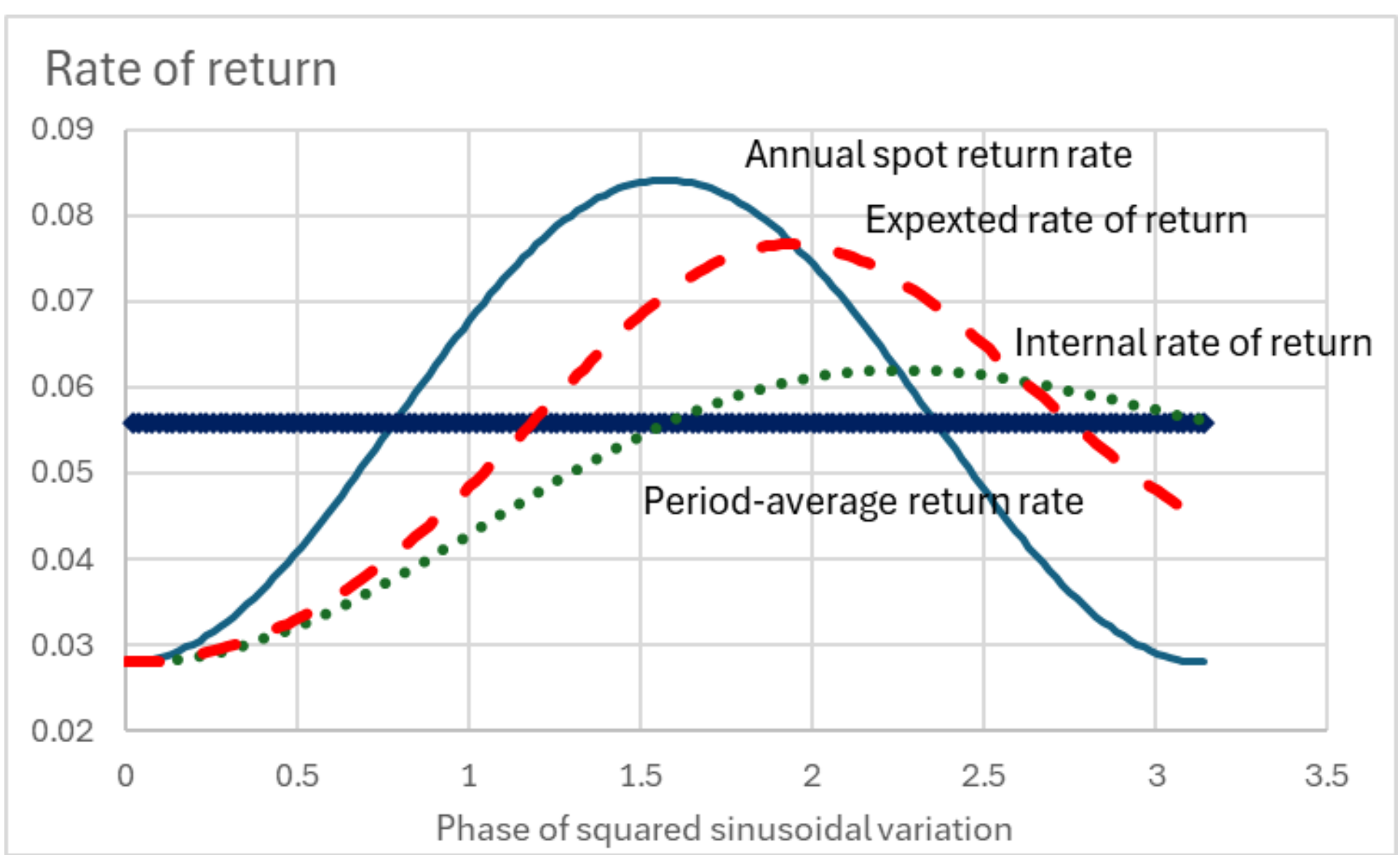

Figure 1. Annual spot return rate according to Eq. (11) with parameter value $a$=0.5, expected value of the rate of return according to Eq. (3), internal rate of return according to Eq. (12), and the reference case $r(t) = \langle\langle r \rangle\rangle$ from Eq. (11).

### 2.3. Net present value and leverage effect

It is of interest to consider the outcome of a net present value analysis in the present example problem. As a modification of Eqs. (7) and (12), the net present value of all future cash flows is

$$NPV = K(0)\frac{e^{\tau(\langle\langle r \rangle\rangle - q)} - 1}{1 - e^{-q\tau}} \qquad (13),$$

where q is a discount rate (as in Eq. (7)), and the denominator corresponds to the contribution of further rotations in the future [1, 52, 53]. The discount rate is supposed to reflect market interest rates, and it is used to match achievable present value to a utility-indifference curve of different temporal consumption patterns [2, 3, 4, 11]. Figure 2 shows the effect of the discount rate on the net present value, as a modification of Fig. 1. The discount rate is given as multiples of the time-average value of the spot return rate on capital $\langle\langle r \rangle\rangle$ within a full cycle $\frac{\Gamma}{\Gamma}\pi$ of the squared periodic function. The discount rate affects not only the level of $NPV$ but also the rotation cycle length where the $NPV$ becomes maximized (Fig 2).

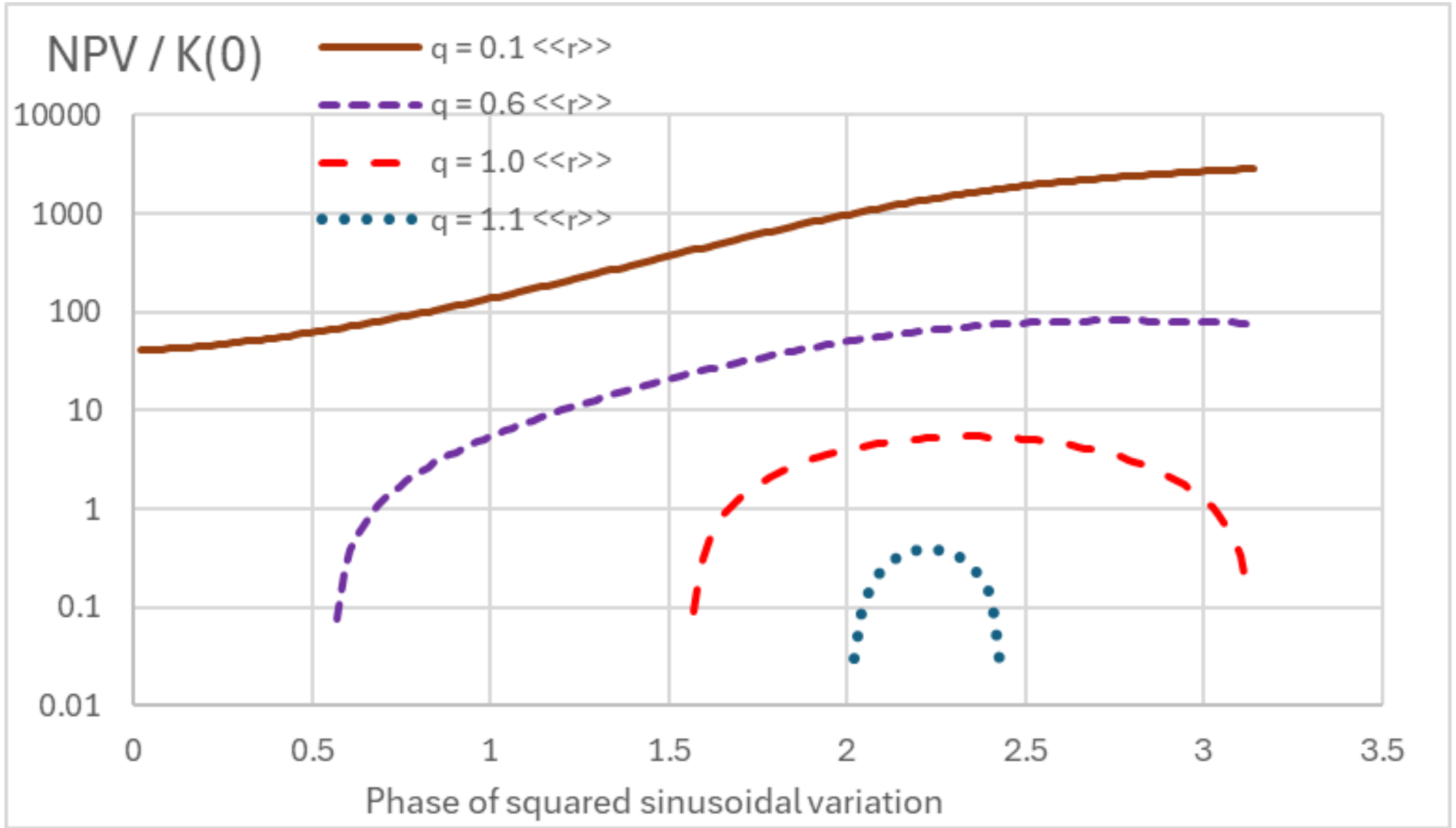

Figure 2. Net present value of future cash events, according to Eq. (13), normalized by the initial investment $K(0)$. The discount rate q is given in different multiples of $\langle\langle r \rangle\rangle$ (Eq. (11)).

It is found from Fig. 2 that the *NPV* varies very much as a function of the discount rate. In general, utilization of the *NPV* for consumption requires that any agent can borrow at the market rate of an assumed perfect financial market [2, 3, 4, 11]. However, there are cases where the discount rate can be derived from the utility function [4]. As the translation along the time axis in terms of discounting is considered financially invariant, and there is no return rate apart from the discount rate, the *NPV*-approach is apparently unable to discuss any leverage effect.

There is a leverage effect. The leverage ratio $L$ is here defined as

$$L+1=\frac{K}{E} \qquad (14),$$

where $K$ is capitalization, and $E$ is equity. Increasing capitalization increases leverage if equity does not change. The capitalization within the productive system may approach zero, in which case the leverage coefficient reaches negative unity. Leverage may also be changed by increasing or decreasing the amount of equity.

Then, the leverage effect on the return rate on capital can be written as

$$\langle RROE \rangle = \langle s \rangle + L\left(\langle s \rangle - u\right) \qquad (15),$$

where $\langle RROE \rangle$ is the expected value of the return rate on equity, and $u$ is a market interest rate [cf. 30]. The rotation cycle length apparently enters Eq. (15) only through $\langle s \rangle$ (or $\langle RROC \rangle$). $\langle RROE \rangle$ can now be plotted as a function of $L$ and $u$; in Figure 3, $L$=1, and the market interest rate is given in terms of multiples of $\langle \langle r \rangle \rangle$. It is found that whereas the $\langle RROE \rangle$ depends on the market interest rate, the rotation cycle length maximizing $\langle RROE \rangle$ indeed is independent of the market interest level. For the market interest rates appearing in Fig. 3, the leveraging increases the return on equity in the vicinity of the cycle length close to the maximum of $\langle s \rangle$. On the other hand, in rotation cycle lengths where $\langle s \rangle$ is less than $u$, the leveraging reduces the $\langle RROE \rangle$. It is here worth noting that seminal research has stated the magnitude of productive investments depend on the market interest rates [7].

As mentioned above, negative leverage is also possible; the range of the leverage is from negative unity to infinity. Negative leverage maximizes the $\langle RROE \rangle$ if the market rate is greater than the expected value of return on capital in the periodic growth process $\langle s \rangle$ (Eq. (3)). Negative leverage corresponds to lending instead of borrowing. The leverage ratio approaching negative unity corresponds to the lending of all available capital, resulting in zero capital allocation to the periodic growth process.

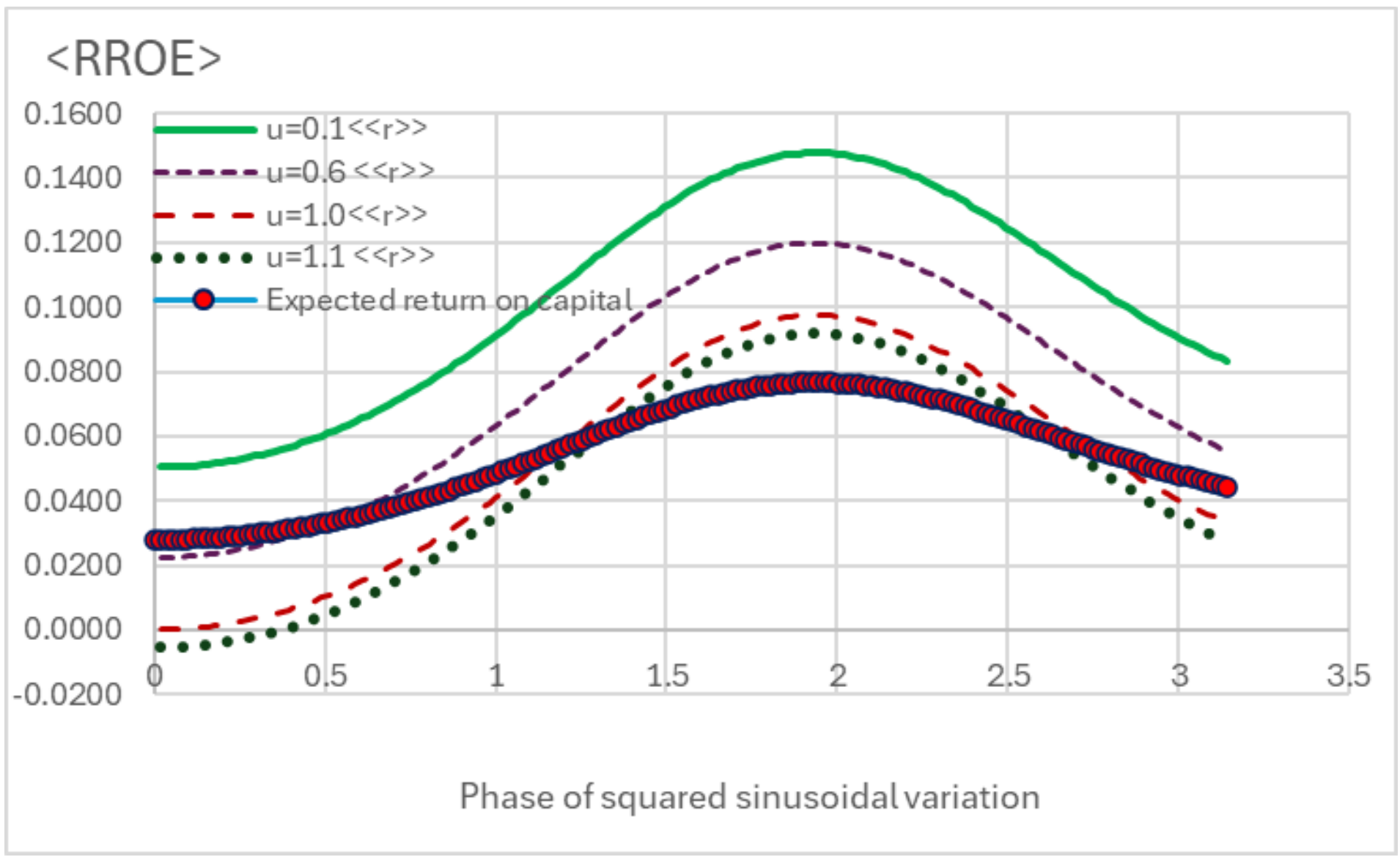

Figure 3. Expected value of the rate of return according to Eq. (3), and leveraged (*L*=1) rate of return on equity according to Eq. (15) with four different interest rates, expressed as multiples of $\langle\langle r \rangle\rangle$ (Eq. (11)).

Another question is, how the leverage effect could be applied in the context of the internal rate of return. Firstly, Eq. (15) could be technically used, simply by replacing $\langle s \rangle$ by the internal rate of return $o$ from Eq. (6) or (12). However, such a solution would not be correct since the *IRR* does not correspond to the expected value of the rate of return on capital (Fig. 2). Alternatively, one could write another *IRR*-type equation for a leveraged project as

$$\left[(1+L)e^{<r>\tau} - Le^{u\tau}\right]e^{-\Omega\tau} - 1 = 0 \qquad (16),$$

where $\Omega$ is a discount rate forcing the initial value of the leveraged project to zero. This discount rate is not internal, since it depends on the market interest rate $u$. The leverage effect is readily solvable from Eq. (16) as

$$\Omega - \langle r \rangle = \frac{1}{\tau}\ln\left[1 + L\left(1 - e^{-\tau[\langle r \rangle - u]}\right)\right] \qquad (17).$$

Eq. (16) has been written under the boundary condition that compound loan interest is paid along with the loan itself at the end of the rotation. Another Equation can be written for the case of interest payments during the growth period; such an Equation is resolvable at least numerically.

A few asymptotic values of Eq. (17) are indeed correct. However, there is no proof that it would produce the leverage effect correctly, including the effect of the market interest rate. There rather is proof that it does not: Eq. (15) and Fig. (2) show that the market interest rate does not contribute to the rotation cycle length maximizing the return rate on equity ( $\langle RROE \rangle$). Eq. (17) conflicts with that result, as the rotation time interacts with the market interest rate in the determination of the leverage effect.

Eq. (16) inspires an eventual application to the leverage effect on *NPV*. Indeed, it is possible to write a leveraged net present value as

$$NPV(L) = K(0)\frac{\left[(1+L)e^{\tau\langle r \rangle} - Le^{\tau u}\right]e^{-\tau d} - 1}{1 - e^{-d\tau}} \qquad (18).$$

Equation (18) contains three financial time rates: the time-average return rate on capital $\langle r \rangle$, the market interest rate *u*, and the discount rate *d*. As the first two first rates can be determined from the production process and from the market, the discount rate remains arbitrary. *NPV(L)* on the left-hand side being a priori unknown, the Equation is then not solvable. However, it has been postulated that the discount rate must equal a market interest rate [4]. Then, Eq. (18) becomes solvable, and the leveraged *NPV* in relation to unleveraged *NPV* becomes

$$\frac{NPV(u=d,L)}{NPV(u=d,L=0)}=\frac{\left[\left(1+L\right)e^{\tau\langle r\rangle}-Le^{\tau u}\right]e^{-\tau d}-1}{e^{\tau\left(\langle r\rangle-d\right)}-1}=1+L \qquad (19).$$

Eq. (19) contains a remarkable result. Yes, indeed, it does describe the effect of the leverage coefficient reasonably. But the leverage effect does not depend on the market interest rate! The latter finding is in serious conflict with Eq. (15) and indicates that the *NPV*-approach seriously fails in its description of the leverage effect. It also is worth noting that leverage coefficient of negative unity renders the net present value to zero. In other words, investment in interest-bearing instruments produces no value.

Even if the utilization of the maximized *NPV* requires borrowing at the discount rate [4], or $u=d$, a more general expansion of Eq. (18) with the help of Eq. (19) might be of interest.

$$NPV(L)=K(0)\frac{\left[\left(1+L\right)e^{\tau\langle r\rangle}-Le^{\tau u}\right]e^{-\tau d}-1}{1-e^{-d\tau}}=NPV\left[1+L\frac{e^{\tau\langle r\rangle}-e^{\tau u}}{e^{\tau\langle r\rangle}-e^{\tau d}}\right] \qquad (20).$$

A few features of Eq. (20) appear logical. The unleveraged *NPV* is positive if $\langle r \rangle > d$. Then, a positive leverage coefficient makes a positive leverage effect if $\langle r \rangle > u$, and negative leverage effect if $\langle r \rangle < u$. The unleveraged *NPV* is negative if $\langle r \rangle < d$. Then, a positive leverage coefficient makes a negative leverage effect if $\langle r \rangle > u$, and positive leverage effect if $\langle r \rangle < u$. However, with negative unleveraged *NPV*, negative leverage effect increases the value of the leveraged *NPV*. Interestingly, the leverage effect goes to zero if $\langle r \rangle = u$, and to infinity if $\langle r \rangle = d$. The effect of the interest rate vanishes if $u = d$.

With negative leverage coefficient *L*, most of the above becomes reversed. For a positive unleveraged *NPV*, there is a negative leverage effect if $\langle r \rangle > u$, and positive leverage effect if $\langle r \rangle < u$. In other words, the higher the interest, the greater the *NPV*. This is understandable since negative leverage corresponds to lending, instead of borrowing. For a negative unleveraged *NPV*, there is a negative leverage effect if $\langle r \rangle > u$, and positive leverage effect if $\langle r \rangle < u$. Again, with negative unleveraged *NPV*, negative leverage effect increases the value of the leveraged *NPV*. Again, the effect of the interest rate vanishes if $u = d$, but if the leverage ratio simultaneously reaches negative unity, the *NPV* goes to zero. This is apparently controversial since the leverage ratio of negative unity means that all equity is invested in interest-bearing instruments. However, the equality of the interest rate *u* and the discount rate q means that interest-bearing instruments produce no value.

## 3. Discussion

The results above contain three apparently nontrivial findings. Firstly, the net present value (*NPV*) in Figure 2 is not produced based on internal features of the production process, but strongly depends on the discount rate. In contrast, both the internal rate of return (*IRR*) and the expected value of the rate of return on capital ($\langle RROC \rangle$ or $\langle s \rangle$) are produced independently of consumption preferences. Then, the optimization of the production process does not depend on market interest rates. It is however worth noting that the magnitude of productive investments within a society depends on the availability of financing [7].

This first result is not new, even if it is possibly not very widely known. It has been shown in the literature that a series of investments made for the purpose of accumulating capital value should be evaluated by the accumulation rate of returns, rather than discussing consumption utility [7, 22, 23, 24, 17, 25, 11, 47]. Surprisingly, regarding this result, erroneous statements have appeared in some esteemed research papers [54, 53, 4].

The second nontrivial result appears to be novel. The internal rate of return (*IRR*) being determined on cash basis, it depends only on the amount and timing of cash events; is path-independent regarding growth-induced changes in capitalization. On the other hand, the expected value of the rate of return on capital ($\langle RROC \rangle$ or $\langle s \rangle$), produced as the ratio the expected value of profit rate to that of the capitalization (Eq. (3)), it depends on the schedule of changes in capitalization. In other words, it is dependent on the capitalization path. Correspondingly, the expected value of the rate of return on capital ($\langle RROC \rangle$ or $\langle s \rangle$) very much differs from the internal rate of return (*IRR*) in Figure 1, where the spot rate of return depends on the phase within the growth cycle according to Eq. (11).

The internal rate of return (*IRR*) depends on the amount and timing of cash events. Correspondingly, it does depend on investments and divestments. In Figure 1, the *IRR* would relate to the capitalization path if there were intermediate investments or divestments. On the other hand, on an accrual basis, the profit rate is only indirectly affected by investments or divestments – their direct effect on Eq. (3) occurs through the denominator.

The third nontrivial result relates to the leverage effect. The rate of return on equity depends on the leverage as given in Eq. (15) and Fig. 3. In analogy with the internal rate of return, a discount rate can be resolved that renders the initial value of a leveraged project to zero (Eq. (17)). Such a discount rate, however, is not internal, as it depends on the market interest rate. Such a postulation (Eq. (17)) corresponds to a few limiting solutions correctly but is not proven to correctly reflect the rate of wealth accumulation. Eq. (17) conflicts with Eq. (15) and Fig. 2, as the rotation time interacts with the market interest rate in the determination of the leverage effect.

Then, leverage was introduced into the net present value in Eq. (18), from which the leverage effect was resolved in Eq. (19). Surprisingly, the market interest rate was absent from the leverage effect. This is clearly an incorrect result, and obviously disqualifies the *NPV* from financial considerations including leverage. A possible reason for the absence of the market interest rate is that it was postulated to be equal to the discount rate, and temporal translation in terms of discounting is assumed financially invariant. On the other hand, the equality of the discount rate to the borrowing interest rate is needed for the matching of achievable present value to a utility-indifference curve [2, 3, 4, 11]. Such an explanation appears to be in concert with the fact that a leverage coefficient of negative unity renders the net present value to zero; investments in interest-bearing instruments cannot produce any value (Eqs. (18) and (19)).

It is worth noting that in Fig. 1, the phase of the rotation where the expected value of the rate of return on capital (*RROI* or $\langle s \rangle$) is maximized differs from the phase where the internal rate of return (*IRR*) is maximized. Correspondingly, the two criteria result in different procedures in the management of the periodic growth process. As Eq. (12) indicates that the internal rate of return $o$ equals the time-average value of spot return rates $\langle r \rangle$, the two terms unify when any spot return rate value $r$ equals $\langle r \rangle$. Within the *ansatz* given in Eq. (11), this would correspond to the parameter $a$ reaching the value of unity.

Is there possibly a reason to discuss the intermediate capitalization within a production process, as is done in the case of Eq. (3)? In other words, are there some benefits of an accrual basis rather than cash basis in financial considerations? Indeed, there are several reasons. Firstly, the production process may be developed to become less capital demanding. Consequently, released capital can be invested to produce interest or capital gain. Secondly, there often is a possibility to divest productive assets. One can expect the prospective income from divesting to relate to capitalization. Then, the opportunity cost to be considered is the lower the lower the intermediate capitalization. Thirdly, capital values may be used as collaterals for loans, even in the absence of timely cash flows.

Fourthly, in production facilities involving growth processes, there often are production sites at different stages of development. Then, Eq. (2) for the expected value of capitalization within the facility can be rewritten as

$$\langle K \rangle = \int_{b}^{b+\tau} K \; p(a) \; da \qquad (20),$$

where $p(a)$ is the probability density of site ages, which may or may not evolve in time [55]. The expected value of capitalization for the entire facility then directly contributes to the rate of return according to Eq. (3), and is subject to streamlining efforts, as well as opportunity cost considerations. Further, by substituting Eq. (5) into Eq. (3), one finds that the expected value of the return rate on capital is weighed by the capitalization on the production sites. In case capitalization varies, the result differs from the area-average of local momentary capital return rates. In the special case of uniform age distribution, or constant $p(a)$, the area-average value would equal the time-average, or the internal rate of return $o = \langle r \rangle$. This shows that the *IRR* is not representative of the capital return on the estate.

It is worth noting that not all sites of periodic production need to have the same period duration. Eq. (21) then must be understood as an integral over a complete period for the entire facility, the period duration corresponding to a common multiple of the production site period durations. Provided the periodic boundary condition is satisfied for the entire facility, the probability density $p(a)$ does not evolve over time.

It was stated above that when capital appreciation is aspired, the rotation cycle length maximizing the expected value of *RROE* is independent of the market interest level. This may not be the entire truth. There obviously is no first-order effect of the market interest on the rotation length maximizing the expected values of *RROC* and *RROE*. However, such return rates depend on prices, as well as on productivity. Prices may be affected by market interests. Also, productivity may depend on interest rates through an eventual effect on investments.

Astonishingly, no references have been found in the literature regarding the path-dependency of the expected value of the return rate on capital on growth processes. The application of the return rate on capital in such processes has been applied only in a handful of publications [56, 57, 58, 59]. This is astonishing since it appears this measure apparently is the most fundamental single measure of financial performance [28, 29, 30, 40], and as shown in this paper, significantly differs from competing measures like the internal rate of return. An obvious explanation for the fundamental role of the rate of return on capital in Eq. (3) is that it is derived from the accounting measures of profit rate and capitalization, unlike the other measures discussed in Eqs. (6) and (7).

In investment portfolio management, capital return is often balanced with risk, as asset types vary in both measures. It is worth asking how return and risk are related in periodic growth processes. Obviously, increasing leverage may increase the return on equity (Eq. (15)), but not without risk [23]. The relation of the return on capital to risk is far less obvious. In optimization problems, the risk of not meeting constraints can be assessed [60]. No constraints have been discussed in this paper, but they probably will appear in management applications.

This paper has applied a few financial procedures to simple systems. In real life, not all systems are that simple. However, any procedure yielding biased results for a simple system is shown not to be useful. On the other hand, procedures working apparently well for simple systems do not always perform well in more demanding applications.

## Acknowledgements

The author declares that no competing interests exist.
This work was partially funded by Niemi foundation. The funder had no role in study design, data collection and analysis, decision to publish, or preparation of the manuscript.

## References

1. Faustmann M. 1849. Berechnung des Wertes welchen Waldboden sowie noch nicht haubare Holzbestande fur die Waldwirtschaft besitzen. Allg Forst- und Jagdz, Dec 1849, 440–455. On the determination of the value which forestland and immature stands pose for forestry. Reprinted in Journal of Forest Economics 1, 7–44 (1995).
2. Fisher I. 1907. The rate of Interest. Macmillan Company, NY. 442 p.
3. Fisher I, 1930. The theory of Interest. Macmillan Company, NY. 566 p.
4. Hirshleifer J, 1958. On the Theory of Optimal Investment Decision. Journal of Political Economy 66, 329-352.
5. Böhm-Bawerk E von. 1889. The Positive Theory of Capital. trans. William A. Smart, London: Macmillan and Co. 1891.
6. Böhm-Bawerk E Von. 1851-1914. Kapital und Kapitalzins. Positive Theorie des Kapitales. Jena, Fischer, 1921.
7. Keynes JM. 1936. The General Theory of Employment, Interest, and Money. Palgrave Macmillan, London. 472 p.
8. Wright JF. 1959. The Marginal Efficiency of Capital. The Economic Journal 69(276), 813–816. https://doi.org/10.2307/2227693
9. Biondi Y, Marzo G, Decision Making Using Behavioral Finance for Capital Budgeting (October 17, 2010). CAPITAL BUDGETING VALUATION: FINANCIAL ANALYSIS FOR TODAY'S INVESTMENT DECISIONS, K. Baker & P. English, eds., Wiley, 2011, Available at SSRN: https://ssrn.com/abstract=1700154
10. Simon HA, Rational Decision Making in Business Organizations. The American Economic Review, vol. 69, no. 4, 1979, pp. 493–513. JSTOR, http://www.jstor.org/stable/1808698. Accessed 9 Dec. 2024.

11. Dorfman R. 1981. The meaning of internal rates of return. The Journal of Finance, 36(5), 1011-1021.
12. Osborne MJ. 2010a. A resolution to the NPV–IRR debate?, The Quarterly Review of Economics and Finance 50, 234-239. https://doi.org/10.1016/j.qref.2010.01.002.
13. Osborne MJ. 2010b. On the Meaning of Internal Rates of Return and Why an Internal Rate of Return is Not an Investment Criterion. http://dx.doi.org/10.2139/ssrn.1634819
14. Ramsey JB. 1970. The Marginal Efficiency of Capital, the Internal Rate of Return, and Net Present Value: An Analysis of Investment Criteria. Journal of Political Economy 78, 1017-1027.
15. Biondi Y. 2006. The double emergence of the Modified Internal Rate of Return: The neglected financial work of Duvillard (1755 – 1832) in a comparative perspective. The European Journal of the History of Economic Thought, 13(3), 311–335. https://doi.org/10.1080/09672560600875281
16. Hildreth C. 1946. A Note on Maximization Criteria. Quarterly Journal of Economics 61, 156-164.
17. Chipman JS. 1977. A Renewal Model of Economic Growth: The Continuous Case. Econometrica 45, 295-316.
18. Galenson W, Leibenstein H. 1955 Investment Criteria, Productivity, and Economic Development. Quarterly Journal of Economics 69, 343-370.
19. Eckstein O. 1957. Investment Criteria for Economic Development and the Theory of Intertemporal Welfare Economics." Quarterly Journal of Economics 71, 56-85.
20. Marglin SA. 1963a. The Social Rate of Discount and the Optimal Rate of Investment. Quarterly Journal of Economics 77, 95-111.
21. Marglin SA. 1963b. Opportunity Costs of Public Investment. Quarterly Journal of Economics 77, 274-289.
22. Lutz F, Lutz V. 1951. The Theory of Investment of the Firm. Princeton University Press, Princeton, N.J.
23. Boulding KE. 1955. Economic Analysis. Harper & Row 1941, 3rd ed. 1955.
24. Chipman JS. 1972. Renewal Model of Economic Growth: The Discrete Case. In “Mathematical Topics in Economic Theory and Computation”, R. H. Day and S. M. Robinson, eds., SIAM Publications, Philadelphia.
25. Miller RK. 1975. A System of Renewal Equations. SIAM Journal on Applied Mathematics 29, 20-34. https://www.jstor.org/stable/2100201
26. Biondi Y. 2024. "Accounting and Finance: Complementarity and Divergence" Accounting, Economics, and Law: A Convivium, vol. 14(3), 329-337. https://doi.org/10.1515/ael-2023-0132
27. Brealey RA, Myers SC, Allen F. 2011. Principles of Corporate Finance (Tenth ed.), New York, NY: McGraw-Hill Irwin.
28. Danaher PJ, Rust RT. 1996. Determining the optimal return on investment for an advertising campaign. European Journal of Operational Research 95(3), 511-521.
29. Li J, Min KJ, Otake T, Van Voorhis T. 2008. Inventory and investment in setup and quality operations under Return On Investment maximization. European Journal of Operational Research, 185(2), 593-605.
30. Magni CA. 2021a. Internal rates of return and shareholder value creation, The Engineering Economist 66, 279-302. DOI: 10.1080/0013791X.2020.1867679
31. Manz P. 1986. Forestry economics in the steady state: the contribution of J. H. von Thünen. History of Political Economy 18, 281-290.
32. Myung Y-S, Kim H, Tcha D. 1997. A bi-objective uncapacitated facility location problem. European Journal of Operational Research 100(3), 608-616.

33. Helmedag F. 2018. From 1849 back to 1788: reconciling the Faustmann formula with the principle of maximum sustainable yield. Eur J Forest Res 137, 301–306. https://doi.org/10.1007/s10342-018-1101-8
34. Chang SJ, Chen Y, Zhang F. 2020. Debunking the forest rent model fallacy in a fully regulated forest. Eur J Forest Res 139, 145–150. https://doi.org/10.1007/s10342-019-01240-z
35. Magni CA. Marchioni A. 2020. Average rates of return, working capital, and NPV-consistency in project appraisal: A sensitivity analysis approach. International Journal of Production Economics 229, 107769. https://doi.org/10.1016/j.ijpe.2020.107769.
36. Hazen G, Magni CA. 2021. Average internal rate of return for risky projects. The Engineering Economist 66, 90-120. DOI: 10.1080/0013791X.2021.1894284
37. Vernimmen P, Le Fur Y, Dallochio M, Salvi A, Quiry P. 2017. Return on Capital Employed and Return on Equity. In “Corporate Finance” (eds P. Vernimmen, Y. Le Fur, M. Dallochio, A. Salvi and P. Quiry). https://doi.org/10.1002/9781119424444.ch13
38. Gomme P, Ravikumar B, Rupert P. 2017. Return to Capital in a Real Business Cycle Model. Federal Reserve Bank of St. Louis Review, Fourth Quarter 2017, pp. 337-50. https://doi.org/10.20955/r.2017.337-350
39. Milano G. 2010. Don't Be Too Preoccupied with Return on Capital (May 19, 2010). https://fortuna-advisors.com/dont-be-too-preoccupied-with-return-on-capital/
40. Vanha-Perttula K, Purola M. 2023. Return on capital (ROE, ROI, ROIC, RONIC), Kaisa V, Miika P, Inderes. https://www.inderes.fi/articles/return-on-capital-roe-roi-roic-ronic Retrieved Feb 10, 2024.
41. Tram Le TH, Huong Le TX. 2023. Return on Invested Capital, Return on Investment, a Measure of the Profitability of Invested Capital, Research Evidence at Song Hong Garment Joint Stock Company. Int. J. Multidisclipnary Research and Analysis 6, 3856-3861. DOI: 10.47191/ijmra/v6-i8-60
42. Hazen GB. 2003. A new perspective on multiple internal rates of return. Engrg. Economist 48(1), 31–51.
43. Hazen GB, 2009. An Extension of the internal rate of return to stochastic cash flows. Management Science 55(6), 1030–1034. https://doi.org/10.1287/mnsc.1080.0989
44. Magni CA. 2021b. Economic profitability and (non)additivity of residual income. Ann Finance 17, 471–499. https://doi.org/10.1007/s10436-021-00388-2
45. Speidel G. 1967. Forstliche Betreibswirtschaftslehre. 2nd Edition 1984, 226 p. Verlag Paul Parey, Hamburg. (in German).
46. Speidel G. 1972. Planung in Forstbetrieb. 2nd Edition. Verlag Paul Parey, Hamburg, 270 p. (in German).
47. Kärenlampi PP. 2019a. Wealth accumulation in rotation forestry – Failure of the net present value optimization? PLoS ONE 14(10), e0222918. https://doi.org/10.1371/journal.
48. Kärenlampi PP. 2019b. The Effect of Capitalization on Financial Return in Periodic Growth. Heliyon 5(10), e02728. https://doi.org/10.1016/j.heliyon.2019.e02728
49. Hu X, Jiang C, Wang H, Jiang C, Liu J, Zang Y, Li S, Wang Y, Bai Y. 2022. A Comparison of Soil C, N, and P Stoichiometry Characteristics under Different Thinning Intensities in a Subtropical Moso bamboo (Phyllostachys edulis) Forest of China. Forests 13, 1770. https://doi.org/10.3390/f13111770
50. Penman, S. 2024. Accounting for Uncertainty. Accounting, Economics, and Law: A Convivium 14(3), 309-327. https://doi.org/10.1515/ael-2022-0059.
51. Cannaday RF. Colwell PF, Paley H. 1986. Relevant and Irrelevant Internal Rates of Return, The Engineering Economist 32, 17-38. DOI: 10.1080/00137918608902950
52. Pearse PH. 1967. The optimum forest rotation. The Forestry Chronicle 43(2), 178-195. https://doi.org/10.5558/tfc43178-2

53. Samuelson PA. 1976. Economics of forestry in an evolving society. Economic Inquiry 14, 466–492. doi:10.1111/j.1465-7295.1976.tb00437.x
54. Samuelson PA. 1937. Some Aspects of the Pure Theory of Capital. The Quarterly Journal of Economics 51, 469-496. https://doi.org/10.2307/1884837
55. Leslie A. 1966. A review of the concept of the normal forest. Australian Forestry 30(2), 139-147.
56. Kärenlampi PP. 2022a. Two Sets of Initial Conditions on Boreal Forest Carbon Storage Economics. PLOS Clim 1(2), e0000008. https://journals.plos.org/climate/article?id=10.1371/journal.pclm.0000008
57. Kärenlampi PP. 2022b. Capitalization and Capital Return in Boreal Carbon Forestry. Earth 3(1), 204-227. https://doi.org/10.3390/earth3010014
58. Kärenlampi PP. 2023. Microeconomics of Nitrogen Fertilization in Boreal Carbon Forestry. Climate 11, 194. https://doi.org/10.3390/cli11090194
59. Kärenlampi PP. 2024. Two Sets of Boundary Conditions in Cyclical Systems with Goodwill in Capitalization. Foundations 4, 3-13. https://doi.org/10.3390/foundations4010002
60. Eyvindson K, Kangas A. 2016. Integrating risk preferences in forest harvest scheduling. Ann For Sci 73(2), 321–330